\documentclass[reprint,aps,pre,showpacs]{revtex4-1}
\usepackage{graphicx}
\usepackage{amsfonts}
\usepackage{amsmath}
\usepackage{csquotes}
\begin{document}
\preprint{Manuscript }
\title{Modulation of ion-acoustic waves in a nonextensive plasma with two-temperature  electrons}
\author{Shalini}
\email{shal.phy29@gmail.com}
\affiliation{Department of Physics, Guru Nanak Dev University, Amritsar-143005, India}
\author{N. S. Saini}
\email{nssaini@yahoo.com}
\affiliation{Department of Physics, Guru Nanak Dev University, Amritsar-143005, India}
\author{A. P. Misra}
\email {apmisra@gmail.com}
\affiliation{Department of Mathematics, Siksha Bhavana, Visva-Bharati University, Santiniketan-731 235, India}
\begin{abstract}
We study the amplitude modulation of  ion-acoustic wave (IAW) packets in an unmagnetized electron-ion plasma with two-temperature (cool and hot) electrons in the context of the Tsallis' nonextensive statistics. Using the multiple-scale technique, a nonlinear Schr{\"o}dinger (NLS) equation is derived which governs the dynamics of modulated  wave packets. It is shown that in nonextensive plasmas, the IAW envelope is always stable for long-wavelength modes $(k\rightarrow0)$  and  unstable for short-wavelengths with $k \gtrsim1$. However, the envelope can be unstable at an intermediate scale of perturbations with $0<k<1$.  Thus, the modulated IAW packets can propagate in the form of bright envelope solitons or rogons (at small- and medium scale perturbations)  as well as dark envelope solitons (at large scale). The stable and unstable regions are obtained for different values of  temperature and  density ratios as well as the nonextensive parameters $q_c$ and $q_h$ for cool and hot electrons. It is found that the more (less) the population of superthermal cool (hot) electrons, the smaller is the growth rate of instability with cutoffs at smaller wave numbers of modulation.
\end{abstract}
\startpage{1}
\endpage{102}
\maketitle
\section{Introduction}
\par
Over the last few decades, wave dynamics in multicomponent plasmas has been a front area of research for the plasma community. Various types of nonlinear  waves namely, ion-acoustic waves (IAWs), electron-acoustic waves (EAWs) etc. as well as nonlinear coherent structures associated with them namely, solitons, envelope solitons, shocks etc.  have also been studied in different plasma environments. As is known, ion-acoustic waves  are   electrostatic plasma modes, where a population of inertial ions oscillate against a dominant thermalized   background of electrons that provide  necessary restoring force.  The phase velocity of such waves lies between the electron and ion thermal velocities. Plasmas with two groups of electrons (e.g., cool and hot electrons) are very common in laboratories \cite{hershkowitz85, nishida86, hairapetian90} as well as in space environments \cite{temerin82}. In this context, a large number of investigations have been made to study the characteristics of IAWs in two-temperature electron plasmas \cite{jon75, gos76, nis81, say93, ric93}. Such two-temperature plasmas have been  observed by various satellite  missions, e.g., Fast Auroral Snapshot (FAST) at the auroral region \cite{pottelette99}, Viking Satellite \cite{temerin82} and THEMIS mission \cite{ergun98}. Furthermore, several investigations made by  GEOTAIL  and  POLAR  \cite{mcfadden03} in the magnetosphere have also confirmed the existence of such electron populations. \textcite{jon75} reported experimentally the  existence of ion acoustic solitary waves (IASWs) in plasmas with hot and cold electrons.
\textcite{bal12} studied small-amplitude ion-acoustic solitons in a plasma with two-temperature kappa distributed electrons and found that superthermality of electrons played an important role for the modification of solitary structures.
\par
The linear properties of the IAWs have been extensively studied and well understood in multi-component plasmas in presence of hot electrons obeying Maxwellian and non-Maxwellian velocity distributions. When the nonlinear effects are concerned, the formation of ion-acoustic solitons may be possible   due to a delicate balance  between the nonlinearity and the dispersion which also has been anticipated theoretically  via the derivation of  Korteweg-de Vries (KdV) or Zakharov-Kuznetsov (ZK) equations for small-amplitude perturbations \cite{was66, tajiri84}, and via the Sagdeev potential formalism, accounting for the arbitrary amplitude excitations \cite{sagdeev66}. Existence of these nonlinear structures was experimentally confirmed \cite{ikezi70}.
\par
The space observations and laboratory experiments  indicate the presence of particles which obey non-Maxwellian velocity distributions in plasmas. The observations of the voyager  PLS  \cite{sit83, bab93} have also shown the existence of both cool and hot electrons populations (non-Maxwellian) in Saturn's Magnetosphere. One of such distributions as reported by \textcite{ren55} and proposed by \textcite{tsa88} is the Boltzmann-Gibbs-Shannon (BGS) entropy in which  the degree of nonextensivity of the plasma particles is characterized by the entropic index $q$, i.e., the distribution function with $q<1$, compared with the Maxwellian one $(q=1)$ indicates the system with more superthermal particles (superextensivity), whereas the $q$ distribution with $q>1$ is suitable for plasmas containing a large number low-speed particles (subextensivity). However, because of long-range Coulomb force between plasma particles and the presence of many superthermal particles in astrophysical and space environments a $q$-distribution with $q<1$ is strongly suggested for   real plasmas or superthermal plasmas. Such  $q$-nonextensive distribution has been widely used to study  various nonlinear  coherent structures like solitons, double layers or shocks etc.  \cite{car08, tri10, nob11, bai11, tri11, sah12, tai12}.
\par
The propagation of wave packets in a nonlinear, dispersive medium is subjected to the modulation of their wave amplitudes, i.e., a slow variation of the wave packet’s envelope due to the nonlinear self-interaction of the carrier wave modes. The system’s evolution is then governed through the nonlinear Schr{\"o}dinger (NLS) equation and the associated modulational instability (MI).  The latter  in   nonlinear, dispersive media has been a well-known mechanism   for the localization of wave energy. Furthermore, such instability  results into the exponential growth of a small plane wave perturbation which leads to the  amplification of the sidebands, and thereby breaking up the uniform wave into a train of oscillations. Thus, the MI of wave packets in plasmas acts  as a precursor for the formation of bright envelope solitons or highly energetic rogue waves (rogons), otherwise   the dark envelope solitons may be formed.
The amplitude modulation of electrostatic waves in   plasmas have been investigated by a number of authors owing to their importance in space, astrophysical and laboratory plasmas.  A great deal of attention has been given to the $MI$ of IAWs in non-Maxwellian plasmas. To mention few, \textcite{saini08} investigated the amplitude modulation of dust-acoustic waves in dusty plasmas with superthermal ions.  Similar investigations have also been made  of IAWs in plasmas with two-temperature electrons \cite{kourakis03} and kappa-distributed superthermal electrons \cite{sultana14}.  Furthermore, \textcite{bai11} studied the MI of ion-acoustic wave envelopes in a plasma containing non-extensive electrons.
\par
On the other hand, the generation of highly energetic rogue waves (or rogons) in  plasmas has been a topic of important research. Such  waves may also appear in other environments such as oceanic rogue waves \cite{kharif09, eliasson10}, stock market crashes \cite{yan10}, optical rogue waves \cite{montina09, kibler10}, super-fluid helium \cite{ganshin08}, Bose-Einstein condensates \cite{bludov09}, atmospheric physics \cite{stenflo10} and opposing current flow \cite{onorato11}. The rogue waves in plasmas may be generated due to wave-wave interactions \cite{shukla06}, and their evolution can also be described by the NLS equation for modulated wave packets. The  rogue waves were first observed in a ocean \cite{peregrine83}. The space-time localized solution (rational solution) of  the NLS equation was determined and later   recognized as Peregrine solitons (a prototype of rogue waves). The latter were also experimentally observed in different nonlinear dispersive media (e.g., optical fiber, plasma etc.). The possible mechanism to explain these waves include effects such as nonlinear focusing via MI in one space dimension. For some  theoretical description of rogue waves readers are referred to Ref.\citenum{dysthe9} and  some experimental observations to Refs.\citenum{clamond06, vor08, kibler10, chabchoub11, bailung11}.
\par
We, however, mention that the physical mechanism behind the formation of rogue waves is still unclear, however, observations indicate that they have unusually steep, solitary or tightly grouped profiles, which appear like \enquote{walls of water} \cite{shukla06}.
A large number of theoretical and experimental investigations have been reported for the study of rogue waves in various plasma systems \cite{ruderman10, shats10, awady11, abdelsalam11, moslem11, bailung11, shukla12, abdelsalam12, bains14, veldes13, guo13, el-wakil14}.
\textcite{abdelsalam12} studied the nonlinear propagation of solitary and freak waves in a superthermal plasma with positive and negative ions, ion beam and stationary dust particles.
More recently, \textcite{shalini15} have considered the propagation of  DIA rogue waves in  superthermal plasmas and investigated the  combined effects of superthermality and dust concentration on the characteristics of DIA rogue waves.
\par
In this work, we consider the amplitude modulation of IAW packets in  an unmagnetized plasma whose constituents are inertial cold positive ions and two-temperature (cool and hot) electrons obeying Tsallis' nonextensive $q$ distribution. The nonlinear evolution of these waves are governed by a NLS equation which is used to study the MI of a Stokes' wave train to a small longitudinal perturbation.   It is shown that the wave packets with long-wavelength modes $(k\rightarrow0)$ are always stable, wheres modes with $k\gtrsim1$ gives rise the MI of IAWs.
The manuscript is organized as follows: In Sec. \ref{II}, we present the fluid model for IAWs. Using the multiple scale perturbation technique, the expressions for different order harmonic modes along with the NLS equation are presented in Sec. \ref{III}.   Section \ref{IV} describes the criteria for  MI and Sec. \ref{V} presents the evolution of bright and dark envelope solitons as well as rogons. Finally,  Sec. \ref{VI} is devoted to summarize and conclude the results.
\section{Fluid model}\label{II}
\par
We consider the nonlinear propagation of IAWs in a collisionless   unmagnetized plasma consisting of $q$-nonextensive distributed cool and hot electrons (to be denoted, respectively, by the subscripts $c$  and $h$) and  singly charged inertial positive ions (with charge $e$ and mass $m_i$). At equilibrium, the overall charge neutrality condition reads  $n_{h0}/n_{i0}=1-f$, where $f=n_{c0}/n_{i0}$ is the ratio between the cool electron and the ion number densities. The dynamics of ion-acoustic perturbations is governed by the following  set of  normalized fluid equations
\begin{equation}
\frac{\partial n_i}{\partial t}+ \frac{\partial (n_{i}{u_i})}{\partial x}=0,\label{e4}
\end{equation}
\begin{equation}
\frac{\partial u_i}{\partial t}+ u_i\frac{\partial u_i}{\partial x}+\frac{\partial \phi}{\partial x}=0,\label{e5}
\end{equation}
\begin{equation}
\frac{\partial^2 \phi}{\partial x^2}= fn_c+(1-f)n_h-n_i, \label{e-6}
\end{equation}
where, $n_j$ is the number density of $j$-species particles  ($j=c,h,i$, respectively, stand for cool electrons, hot electrons and positive ions) normalized by their  equilibrium value $n_{j0}$ and $u_i$ is the ion-fluid velocity normalized by the ion-acoustic speed  $C_{s}=\sqrt{k_{B}T_{c}/m_{i}}$ with $k_B$ denoting the Boltzmann constant. Also, $\phi$ is the electrostatic wave potential normalized by $k_BT_{c}/e$. Furthermore, the  space and the time variables are  normalized by the Debye length $\lambda_{D}=\left(k_{B}T_{c}/4\pi n_{i0}e^{2}\right)^{1/2}$ and the ion plasma period $\omega_{pi}^{-1}=\left(4\pi n_{i0}e^{2}/m_{i}\right)^{-1/2}$ respectively.    In Eq. \eqref{e6}, the number densities $n_c$ and $n_h$ for cool and hot electrons $(j=c,h)$ are given by the following \emph{q}-nonextensive distributions \citep{tsa88}
\begin{equation}
n_{j}=n_{j0}\left[1+\frac{e\phi (q_{j}-1)}{k_BT_{j}}\right]^{(1/2)+1/(q_j-1)}, \label{nj}
\end{equation}
where  $q_{c}$ $(q_{h})$ is a nonextensive parameter for cool (hot) electrons, measuring deviation from the Maxwellian distribution with $q_j\rightarrow1$.
\par
In the propagation of  small but finite amplitude IAWs, one must have $|e\phi/k_BT_c|\ll1$. Also, as mentioned in the introduction that a $q$ distribution with $q<1$ is strongly suggested for real plasma systems or superthermal plasmas. So, we take $q_{c,h}<1$. Thus, Eq. \eqref{nj} can be  binomially expanded (retaining terms up to $\phi^3$) to obtain from Eq. \eqref{e-6} the following:
\begin{equation}
\frac{\partial^2 \phi}{\partial x^2}= 1+(c_{1}+d_{1})\phi+(c_{2}+d_{2})\phi^{2}+
(c_{3}+d_{3})\phi^{3}-n_i, \label{e6}
\end{equation}
where the coefficients $c_j$ and $d_j$  $(j=1,2,3)$ are all positive and given by
\begin{equation}
c_{1}=\frac{1}{2}f(q_{c}+1),~~  c_{2}=\frac{1}{8}f(q_{c}+1)(3-q_{c}), \nonumber
\end{equation}
\begin{equation}
c_{3}=\frac{1}{48}f(q_{c}+1)(3-q_{c})(5-3q_{c}),~d_{1}=\frac{1}{2}(1-f)\beta(q_{h}+1),\nonumber
\end{equation}
\begin{equation}
d_{2}=\frac{1}{8}(1-f)\beta^{2}(q_{h}+1)(3-q_{h}),\label{e12}
\end{equation}
\begin{equation}
d_{3}=\frac{1}{48}(1-f)\beta^{3}(q_{h}+1)(3-q_{h})(5-3q_{h}). \nonumber
\end{equation}
where  $\beta~(={T_{c}}/{T_{h}})$ is the ratio of cool to hot electron temperatures.
\section{Derivation of NLS equation: Pertubative approach}\label{III}
\par
We employ a multiple-scale perturbation technique \cite{asa69,kou05} to derive the evolution equation of a slowly varying weakly nonlinear wave amplitude of IAW packets in a $q$ nonextensive plasma. Here, we consider $A$ as the state vector $\{A\}(=n_{i}$, $u_{i}$, $\phi$), describing the system's state at a given position $x$ and time $t$. Small deviations will be considered from the equilibrium state $A^{(0)}=(1,0,0)^T$ by taking $A=A^{(0)}+\epsilon A^{(1)}+\epsilon^2 A^{(2)}+\cdots=A^{(0)}+\sum_{n=1}^{\infty} \epsilon^n A^{(n)}$, where $\epsilon\ll1$ is a small positive parameter measuring the weakness of the wave amplitude. The wave amplitude is thus allowed to depend on the stretched (slow) coordinates of space and time as $X_n=\sum_n\epsilon^n x$ and $T_n=\sum_n\epsilon^n t$, respectively, where $n=1,2,3,\ldots$ (viz. $X_1=\epsilon x$, $X_2=\epsilon^2 x$, and so forth; same for time), distinguished from the (fast) carrier variables $x$ ($\equiv X_0$) and $t$ ($\equiv T_0$). All the perturbed states depend on the fast scales via the phase $\theta_1=kx-\omega t$ only ($\omega,~k$ are the wave frequency and wave number respectively), whereas the slow scales only enter the $l$th harmonic amplitude $A_l^{(n)}$. Thus,  $A^n=\sum_{l=-\infty}^{\infty}A_l^{(n)}(X,T)e^{il(kx-\omega t)}$ in which the reality condition $A_{-l}^{(n)}=A_{l}^{(n)\ast}$ is met by all the state variables.
\par
In what follows, we substitute the above expansion into Eqs. (\ref{e4}), \eqref{e5} and (\ref{e6}) and equate coefficients of different powers of $\epsilon$. Thus, equating the coefficients of $\epsilon$ for $n=1$, $l=1$, one obtains
\begin{equation}
-i\omega n_{1}^{(1)}+iku_{1}^{(1)}=0,\label{e16}
\end{equation}
\begin{equation}
-i\omega u_{1}^{(1)}+ik\phi_{1}^{(1)}=0,\label{e17}
\end{equation}
\begin{equation}
n_{1}^{(1)}-(k^{2}+c_{1}+d_{1})\phi_{1}^{(1)}=0.\label{e18}
\end{equation}
The solutions for the first harmonics read
\begin{equation}
n_{1}^{(1)}=(k^{2}+c_{1}+d_{1})\phi_{1}^{(1)},\label{e19}
\end{equation}
\begin{equation}
u_{1}^{(1)}=\frac{\omega}{k}(k^{2}+c_{1}+d_{1})\phi_{1}^{(1)}=\frac{\omega}{k}n_{1}^{(1)}.\label{e20}
\end{equation}
Thus, eliminating $n_{1}^{(1)}$ and $\phi_{1}^{(1)}$ we obtain the following linear dispersion relation for IAWs
\begin{equation}
\omega=k/\sqrt{k^{2}+c_{1}+d_{1}}.\label{e21}
\end{equation}
This clearly shows the dependency of the wave frequency of IAWs on the nonextensive parameters $q_{c}$ and $q_{h}$  as well as the density and temperature ratios  $f$ and $\beta$. It is  seen that the wave frequency increases slowly with $k$, however, it decreases  with increasing values of   $q_j,~f$ and $\beta$.  We note that since $c_{1}$ and $d_1$ are positive  the linear stability is ensured in Eq. (\ref{e21}).
\par
For $n=2$ and $l=1$, we obtain a compatibility condition in the form
\begin{equation}
\frac{\partial \phi_{1}^{(1)}}{\partial T_{1}}+V_{g}\frac{\partial \phi_{1}^{(1)}}{\partial X_{1}}=0,\label{e23}
\end{equation}
where   $V_{g}\equiv {\partial \omega}/{\partial k} $ is the group velocity of the IAW packets given by
\begin{equation}
V_{g}=\frac{c_{1}+d_{1}}{(k^2+c_1+d_1)^{3/2}}.\label{e24}
\end{equation}
Clearly, the group velocity of the wave becomes constant in the long-wavelength limit $k\rightarrow0$. This implies that in  a frame moving with the group velocity $V_g$, the time derivatives of all physical quantities should vanish and thus one can observe a slow variation of the wave amplitude in the moving frame of reference. From Eq. \eqref{e24} we also find that the group velocity decreases not only with the wave number $k$ of the carrier wave, but also with increasing values of the  plasma parameters $q_j~(j=c,h),~f$ and $\beta$.
\par
Next, the expressions for the wave amplitudes corresponding to $n=2,l=1$ are obtained as
\begin{equation}
n_{1}^{(2)}=-2ik\frac{\partial \phi_{1}^{(1)}}{\partial X_{1}}\label{e25}
\end{equation}
\begin{equation}
u_{1}^{(2)}=-i\omega\frac{\partial \phi_{1}^{(1)}}{\partial X_{1}}\label{e26}
\end{equation}
Similarly, for $n=2$ and $l=2$, the evolution equations provide the amplitudes of the second order, second harmonic modes which are   proportional to $\left(\phi_{1}^{(1)}\right)^{2}$. The expressions for them are obtained as:
\begin{equation}
n_{2}^{(2)}=C_{1}^{(22)}\left(\phi_{1}^{(1)}\right)^{2},~~u_{2}^{(2)}
=C_{2}^{(22)}\left(\phi_{1}^{(1)}\right)^{2},\nonumber
\end{equation}
\begin{equation}
\phi_{2}^{(2)}=C_{3}^{(22)}\left(\phi_{1}^{(1)}\right)^{2},\label{e28}
\end{equation}
where the coefficients are
\begin{equation}
C_{1}^{(22)}=(c_{2}+d_{2})+(4k^{2}+c_{1}+d_{1})C_{3}^{(22)},\nonumber
\end{equation}
\begin{equation}
C_{2}^{(22)}=\frac{\omega}{k}\left[C_{1}^{(22)}-(k^{2}+c_{1}+d_{1})^{2}\right],\nonumber
\end{equation}
and
\begin{equation}
C_{3}^{(22)}=-\frac{(c_{2}+d_{2})}{3k^{2}}+\frac{(k^{2}+c_{1}+d_{1})^{2}}{2k^{2}}.\label{e29}
\end{equation}
Now, for the zeroth harmonic modes we consider the expressions for  $n=2$, $l=0$ and $n=3$, $l=0$. Thus, we obtain
\begin{equation}
n_{0}^{(2)}=C_{1}^{(20)}\left(\phi_{1}^{(1)}\right)^{2},~u_{0}^{(2)}=C_{2}^{(20)}\left(\phi_{1}^{(1)}\right)^{2},\nonumber
\end{equation}
\begin{equation}
\phi_{0}^{(2)}=C_{3}^{(20)}\left(\phi_{1}^{(1)}\right)^{2},\label{e30}
\end{equation}
where the coefficients are
\begin{equation}
C_{1}^{(20)}=[(c_{1}+d_{1})C_{3}^{(20)}+2(c_{2}+d_{2})],
\end{equation}
\begin{equation}
C_{2}^{(20)}=-\frac{2\omega}{k}(k^{2}+c_{1}+d_{1})^2+V_{g}C_{1}^{(20)},
\end{equation}

\begin{equation}
C_{3}^{(20)}=\frac{2(c_{2}+d_{2})V_{g}^{2}-(k^{2}+3(c_{1}+d_{1}))}{1-(c_{1}+d_{1})V_{g}^{2}}.\label{e31}
\end{equation}
\par
Finally, for $n=3,~l=1$, we obtain equations for third order, first harmonic modes in which    the coefficients of $\phi_1^{(3)}$ and $\partial \phi^{(2)}_1/\partial X_1$ vanish  by the dispersion relation and the group velocity expression. In the reduced equation, we substitute the expressions for  second order, zeroth harmonic modes. Thus, we obtain an equation in the form
\begin{equation}
i\left[\frac{\partial \phi_{1}^{(1)}}{\partial T_{2}}+V_{g}\frac{\partial \phi_{1}^{(1)}}{\partial X_{2}}\right]+P\frac{\partial^{2}\phi_{1}^{(1)}}{\partial X_{1}^{2}}+Q|\phi_{1}^{(1)}|^{2}\phi_{1}^{(1)}=0,\label{e32}
\end{equation}
where the coefficient of dispersion $P$ and the nonlinearity $Q$ are given by
\begin{equation}
P\equiv\frac{1}{2}\frac{\partial^2\omega}{\partial k^2}=-\frac{3}{2}\frac{k(c_{1}+d_{1})}{(k^2+c_1+d_1)^{5/2}},\label{e33}
\end{equation}
\begin{equation}
\begin{split}
Q=&k(k^2+c_1+d_1)^{-3/2} \\
&\times\left[(c_{2}+d_{2})\left(C_{3}^{(20)}+C_{3}^{(22)}\right)
+\frac{3}2(c_{3}+d_{3})\right]\\
&-\frac{\omega}{2}\left(C_{1}^{(20)}+C_{1}^{(22)}\right)-k\left(C_{2}^{(22)}+C_{2}^{(20)}\right).\label{e34}
\end{split}
\end{equation}
The nature of these coefficients will be discussed in detail in the next section.
\par
The total (first order) solution obtained for the electric potential is
\begin{equation}
\phi\simeq\phi_{1}^{(1)}=\phi(x,t)e^{i(kx-\omega t)}+\phi^\ast(x,t)e^{-i(kx-\omega t)},\label{e37}
\end{equation}
where the variation of $\phi(x,t)$ is assumed to be slower than that of $\exp{[i(kx-\omega t)]}$ ($\epsilon=1$ was formally set here, with the understanding that $|\phi|=|\phi_{1}^{(1)}|\ll1$, i.e., remains small). The compatibility condition is now written in the form of a NLS  equation for   small and slowly varying wave amplitude   $\phi(x,t)$ as
\begin{equation}
i\left(\frac{\partial \phi}{\partial t}+V_{g}\frac{\partial \phi}{\partial x}\right)+P\frac{\partial^{2}\phi}{\partial x^{2}}+Q|\phi|^{2}\phi=0.\label{e38}
\end{equation}
Performing a Galilean transformation, namely $\xi=x-V_gt,~\tau=t$,   Eq. \eqref{e38} can be written in its standard form as
\begin{equation}
i\frac{\partial \phi}{\partial \tau}+P\frac{\partial^{2}\phi}{\partial \xi^{2}}+Q|\phi|^{2}\phi=0\label{e39}
\end{equation}
where dispersive and nonlinear coefficients $P$ and $Q$ respectively are given by Eqs. (\ref{e33}) and (\ref{e34}). Since $P$ and $Q$ are dependent on various physical parameters viz $f$, $\beta$, $q_c$ and the carrier wave number $k$, so it is very important to understand the nature of solitary structures from the solutions of Eq.(\ref{e39}) under the influence of such parameters.
\section{Modulational instability}\label{IV}
We consider  the amplitude modulation of a plane wave solution of  Eq. (\ref{e39}) of the  form $\phi=\phi_0 e^{-i\Omega_0 \tau}$,   where  $\Omega_0=-Q|\phi_0|^2$ is the nonlinear frequency shift  with $\phi_0$ denoting the potential of the wave pump. We then modulate the wave amplitude with a plane wave perturbation with frequency $\Omega$ and wave number $K$ as $\phi=\left(\phi_0+\epsilon \phi_1 e^{iK\xi-i\Omega \tau}+c.c.\right)e^{-i\Omega_0 \tau}$.  One thus obtains the following dispersion relation for the modulated IAW packets.
\begin{equation}
\Omega^2=\left(PK^2\right)^2\left(1-\frac{{K}_c^2} {K^2}\right),\label{e40a}
\end{equation}
where $K_c=\sqrt{2|Q/P|}|\phi_0|$ is the critical value of the wave number of modulation. Thus, the  MI sets in for $K<K_c$ and the wave will be modulated for $PQ>0$, or  more precisely, for $Q<0$ since $P$ is always negative. In this case, the perturbations grow exponentially during the propagation of IAW packets. On the other hand, for $K>K_c$, i.e., for $PQ<0$, or $Q>0$, the IAWs are said to be stable under the modulation. The  growth rate of MI can be obtained from Eq. \eqref{e40a} as
\begin{equation}
\Gamma=|P|K^2\left(\frac{K_c^2}{K^2}-1\right). \label{e43}
\end{equation}
Clearly, the maximum value $\Gamma_{\text{max}}$ of $\Gamma$ is achieved at $K=K_c/\sqrt{2}$ and is given by $\Gamma_{\text{max}}=|Q||\phi_0|^2$.
\par
We note that the coefficients $P,~Q$ and hence the critical value $K_c$  as well as the growth rate $\Gamma$ all depend not only on the plasma parameters $q_j,~f$ and $\beta$, but also on the carrier wave number $k$.  In the range of values of $k$, for which the MI is excluded, i.e., $PQ<0$ or $Q>0$, the perturbations may develop into dark solitons which represent an electric potential envelope hole with finite or vanishing potential at the origin. Such solitons occur in  plasmas under the   effects of self-modulation.  On the other hand, for $PQ>0$, i.e., for  $Q<0$, the excitation of   bright envelope solitons may be possible. Although, envelope solitons   appear  to be analytical coincidence via this model, and the phenomena should be clearly distinguished [linear amplitude stability analysis in one case, nonlinear  theory of partial differential equations (PDE) in the other], it has been postulated in another context \cite{dauxois06} that MI is the first evolutionary stage towards the formation of   envelope solitons.  The latter in the form of  bright type are widely observed in abundance in space plasmas \cite{kou05, has75, saini08}.
\par
 Inspecting on the  expressions for $P$ and $Q$ it is found that $P$ is always negative as $c_1,~d_1>0$, however, $Q$ can be   positive or negative depending on the range of values of $k$ and the parameters $q_j,~f$ and $\beta$. Now, for $k\ll1$,  $P$ can be written as $P\simeq-p_{0}k$, i.e.,  $P$ tends to vanish as  $k\rightarrow0$. This is expected for an ion-acoustic mode where $\omega\sim k$ and $d^2\omega/dk^2=0$.  However, $Q$ increases as $Q\simeq{q_{0}}/{k}$ in the limit $k\rightarrow0$. These are in qualitative agreement with earlier theoretical investigations \cite{saini08, sultana11}. Also, in the limit $k\rightarrow0$, the real quantities $p_{0}$ and $q_{0}$ are, in fact, both positive, given by
\begin{equation}
p_{0}=\frac{3}{2(c_{1}+d_{1})^{3/2}},\label{e35}
\end{equation}
 \begin{equation}
q_{0}=\frac{1}{12}\left(\frac{1}{c_{1}+d_{1}}\right)^{3/2}\left[2(c_{2}+d_{2})-3(c_{1}+d_{1})^{2}\right].\label{e36}
\end{equation}
Thus, we find that the IAW packets with long-wavelength carrier wave modes are always stable under the modulation. In the opposite limit, i.e., for $k\gtrsim1$, we will see from the numerical investigation that the waves are mostly modulationally  unstable.
\par
\begin{figure}[ht]
\includegraphics[height=2in,width=3.3in]{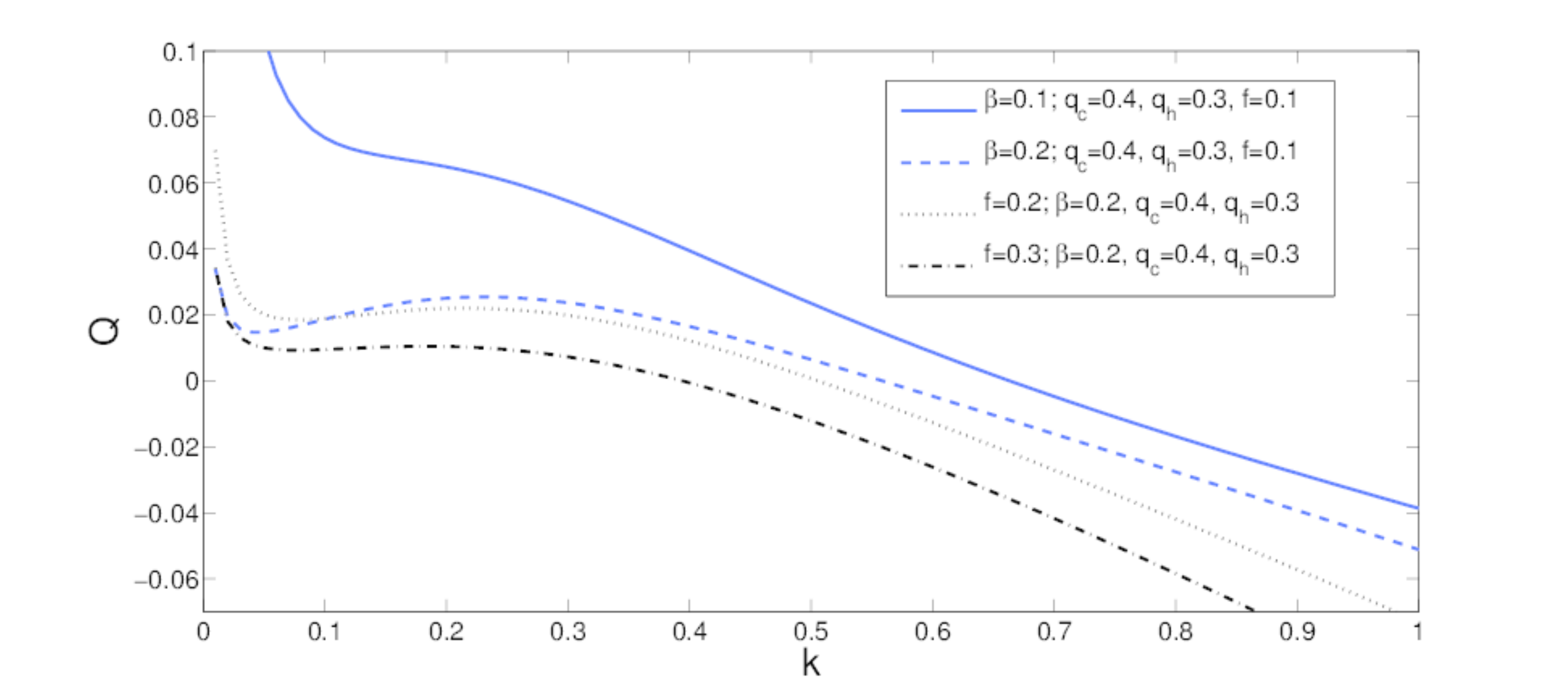}
\label{fig1a}
\includegraphics[height=2in,width=3.3in]{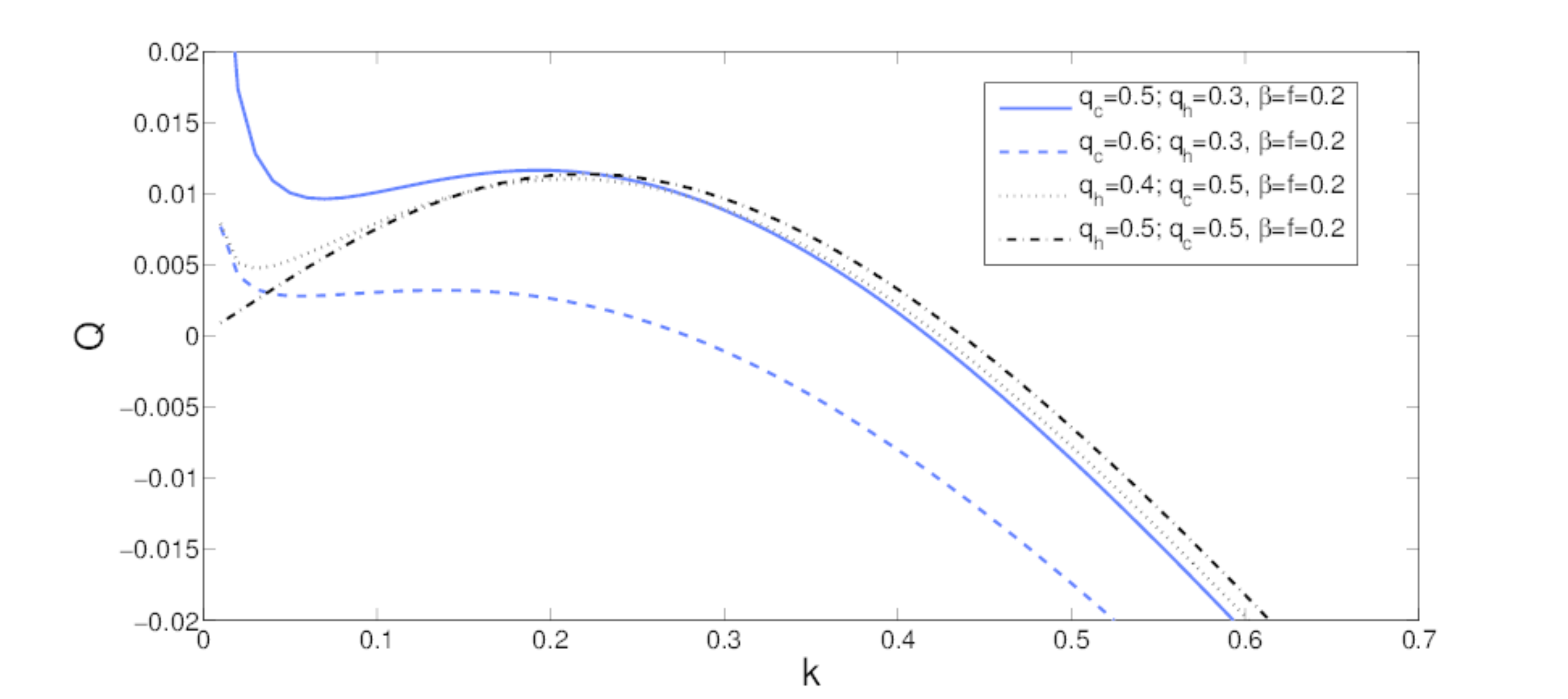}
\label{fig1b}
\caption{Plot of the nonlinear coefficient $Q$ of the NLS equation \eqref{e39} versus  the carrier wave number $k$ is shown for different values of the parameters $\beta,~f,~q_c$ and $q_h$ as in the upper and lower panels.  Since $P$ is always negative, the stable $(PQ<0)$ and unstable $(PQ>0)$ regions are corresponding to $Q>0$ and $Q<0$ respectively.  }
\label{fig1}
\end{figure}

Since $P$ is always negative, the sign of $PQ$, for which the stability and instability of the IAW envelopes can be determined,  solely depends on the sign of $Q$.  So, we numerically investigate the nonlinear coefficient $Q$ for different values of the parameters $q_j,~f$ and $\beta$. Typical behaviors  of $Q$ with respect to $k$  are shown in  Fig. \ref{fig1}.  While the upper panel shows the properties of $Q$ for different values of $\beta$ and $f$, the lower one shows the same with the variations of the nonextensive parameters $q_c$ and $q_h$. It is seen that whatever be the values of the plasma parameters, $Q$ is always positive in the limit  $k\rightarrow0$ (i.e., at long scale perturbations of the carrier modes) as mentioned above, whereas it becomes negative for $k>1$ as well as for some intermediate values of $k$ in $0<k<1$, i.e. at small and medium scale perturbations. It is also evident that $Q$ becomes maximum (hence the maximum growth rate of instability) in the limit $k\rightarrow0$ as mentioned before except for the case where   $q_c,~q_h$ increase  from some positive values to the value $1$ (see the dotted and dash-dotted lines in the lower panel) in which $Q$ can be maximum at an intermediate value  of $k$ in the interval $0<k<1$. Typical behavior of $Q$ (in  the case of $q_c\neq q_h<1$)  shows that initially it drops down from its maximum value in the domain $0<k\lesssim0.1$ and then starts increasing in a small interval of $k\gtrsim0.1$, and again it decreases faster the larger is the wave number $k$ of carrier modes. However, as the values of the nonextensive parameters $q_j$ increase and approach a value $1$ (i.e., one approaches the Maxwellian distribution of electrons) $Q$ initially increases in a small interval $0<k<0.5$, and then it   decreases faster with increasing values of  $k$. From Fig. \ref{fig1}, we also observe that as the parameters   $\beta$ and $f$ get enhanced, the values of $Q$ decrease, and the ranges of values of $k$ for which $Q<0$ increase (See the upper panel). The same are true for increasing values of $q_c$ (See the lower panel), but decreasing values of $q_h$ [except in the range where $k\rightarrow0$ in which the value of $Q$ increases with decreasing values of $q_h$, however, the range of $k$ for which $Q<0$ remains unaltered  (See the dotted and dash-dotted lines the lower panel)].   Thus, from Fig. \ref{fig1} we conclude that the IAW packets with short-wavelength carrier modes $(k>1)$
are always unstable with $Q<0,~P<0$  under the amplitude modulation with plane wave perturbation, it can also be unstable at the medium scale with $k$ lying in $0.5\lesssim k<1$ or more generally in $0<k<1$. In these cases the modulational instability   leads  to the formation of bright envelope solitons or highly energetic rogue waves (rogons).  On the other hand, the IAW packets can be stable ($Q>0,~P<0$) for long-wavelength carrier modes with $k<<1$ leading to  the formation of dark envelope solitons.
\par
\begin{figure}[ht]
\includegraphics[height=2in,width=3.3in]{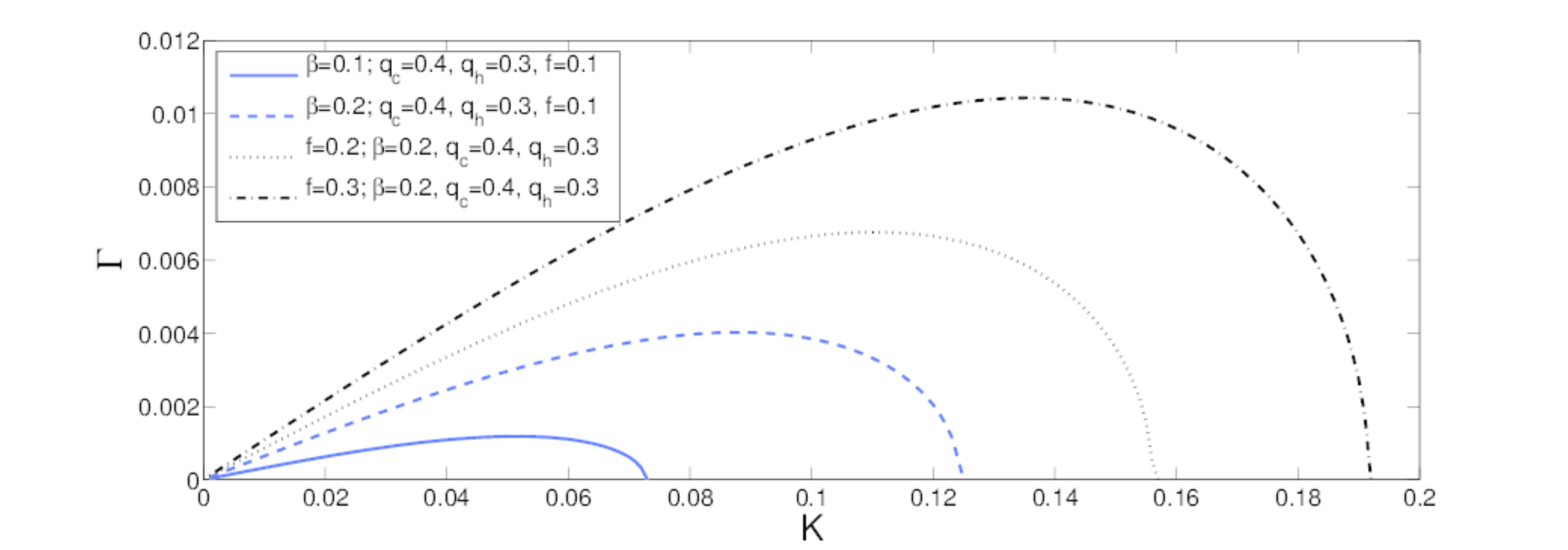}
\label{fig2a}
\includegraphics[height=2in,width=3.3in]{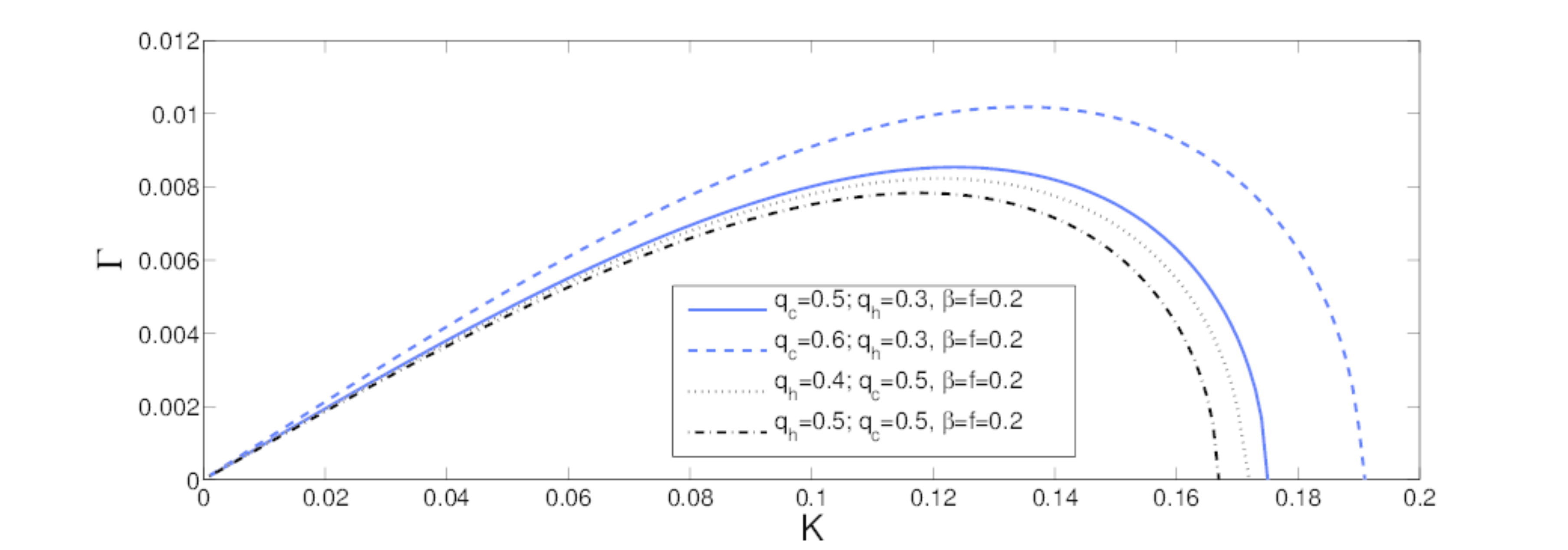}
\label{fig2b}
\caption{The growth rate of instability $\Gamma$ is shown against the wave number of modulation $K$ with the same parameters $\beta,~f$ and $q_j$  as in Fig. \ref{fig1}. The other parameter values are $k=0.7$ and $\phi_0=0.5$. It is seen that the lower (higher) values of $q_c,~f$ and $\beta$ $(q_h)$   are to suppress $\Gamma$ with cutoffs at lower (higher) wave numbers of modulation.  }
\label{fig2}
\end{figure}

In what follows, we   numerically  investigate the growth rate of instability $\Gamma$  for different values of the plasma parameters  $q_j,~f$ and $\beta$ as in Fig. \ref{fig1}. The behaviors of $\Gamma$ are shown in Fig. \ref{fig2}. The critical values $K_c$ of the wave number of modulation $K$ below which the MI sets in can easily  be calculated. It is found that as the parameters $\beta,~f$ and $q_c$ increase, the values of $K_c$ increase, however, the same decreases with increasing values of $\beta_h$. This is expected as $K_c\sim Q/P$.   While the upper panel shows the variation with $\beta$ and $f$, the lower panel exhibits growth rates  for $q_c$ and $q_h$. We find that a small increase of the values of $\beta,~f$ (upper panel) and $q_c$ (lower panel), leads to a significantly higher growth rate of instability with cutoffs at higher values of $K$. However, as $q_h$ increases, the growth rate decreases (but not significantly lower) with cutoff at lower $K$. Thus, we find that the growth rate of instability can be suppressed in plasmas with higher temperature of hot electrons than the cold electrons, the lower concentration of cold electrons than ions and more (less) superthermal cold (hot) electrons.
\section{Envelope solitons and rogons} \label{V}
In the previous section we have  mentioned  that when $PQ>0$, the MI of IAW packets can  give rise to the formation of bright envelope solitons  due to energy localization. In this case, an exact analytic (soliton) solution of  the NLS equation \eqref{e39} can be obtained by considering $\phi=\sqrt{\psi}\exp(i\theta)$, where $\psi$ and $\theta$ are real functions [see for details, e.g., Refs. \citenum{kou05, fedel02, fed02}
as
\begin{equation}
\begin{split}
&\psi=\psi_{b0}~\text{sech}^2\left(\frac{\xi-U\tau}{W_b}\right),\\
&\theta=\frac{1}{2P}\left[U\xi+\left(\Omega_0-\frac{U^2}{2}\right)\tau\right].\label{bright-envelope}
\end{split}
\end{equation}
This solution, in fact, represents a localized pulse traveling at a speed $U$ and oscillating at a frequency  $\Omega_0$ at rest. The width $W_b$  and the constant amplitude $\psi_{b0}$ of the pulse are related to $W_b=\sqrt{2P/Q\psi_{b0}}$. It follows that for a constant amplitude, as the parameters $\beta,~f$ and $q_c$ increase, the width of the envelope soliton decreases, however, the same increases with increasing values of $\beta_h$.
\par
\begin{figure}[ht]
\includegraphics[height=2in,width=3.3in]{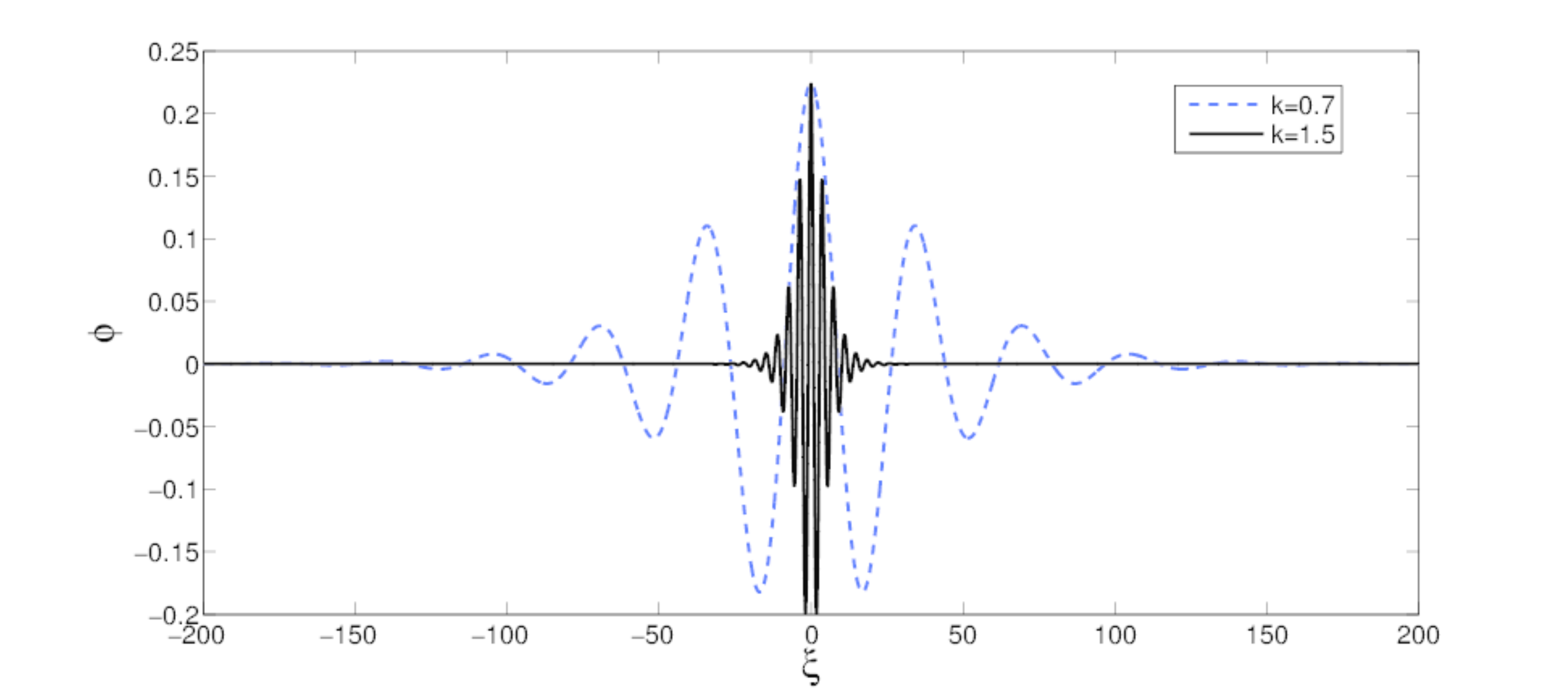}
\label{fig3a}
\includegraphics[height=2in,width=3.3in]{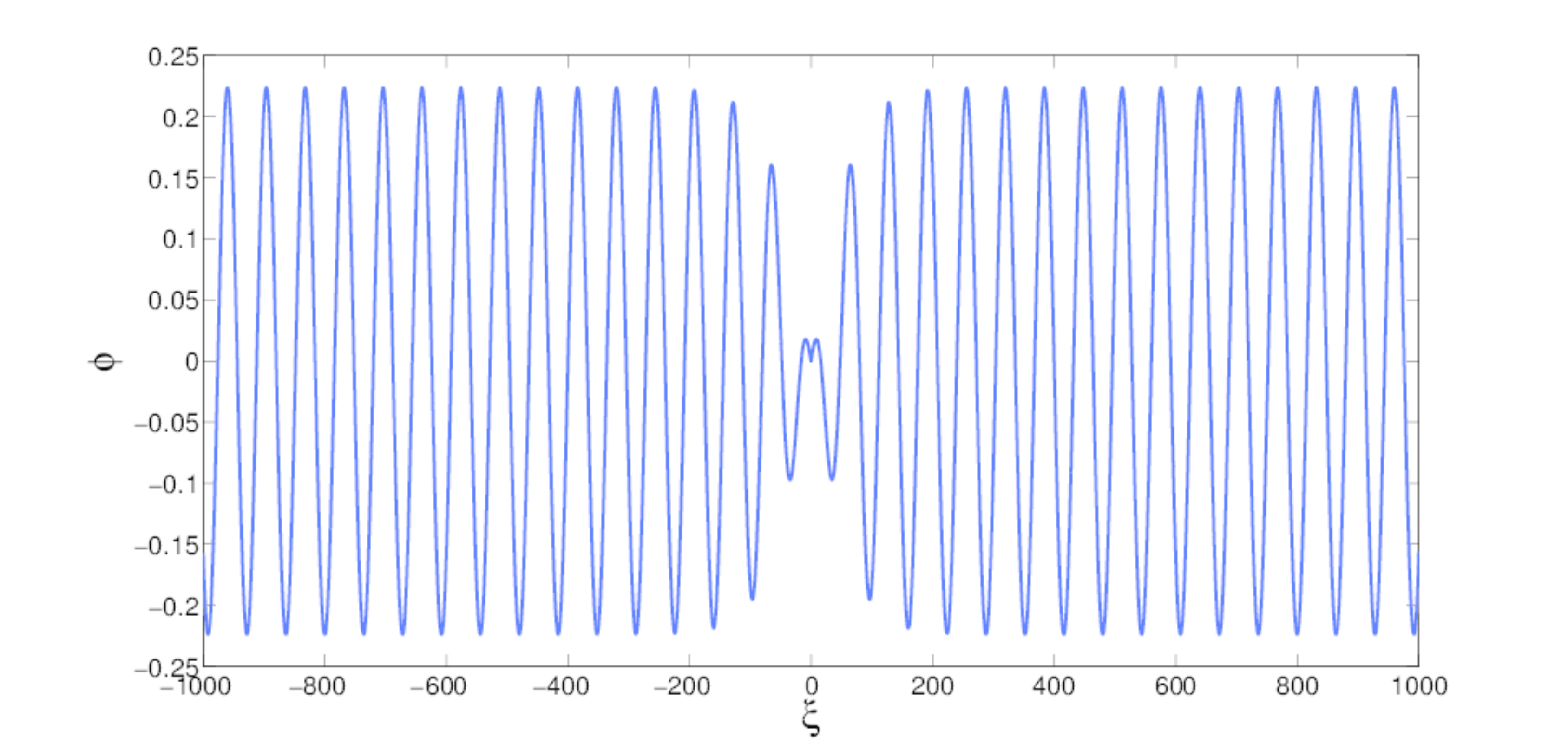}
\label{fig3b}
\caption{Plots of the bright (upper panel) and dark (lower panel) envelope solitons at $\tau=0$ given by Eqs. \eqref{bright-envelope} and \eqref{dark-envelope} respectively. It is seen that while the bright envelope solitons exist both at the small (e.g., $k=1.5$) and medium  (e.g., $k=0.7$) scales, the dark solitons exist only at the large scales (e.g., $k=0.1$). The parameter values of $q_j,~f$ and $\beta$ are the same as for the dotted line in the lower panel of Fig. \ref{fig1}. The other values are taken as $\psi_{b0}=\psi_{d0}=0.05,~U=0.2$ and $\Omega_0=0.5$.   }
\label{fig3}
\end{figure}

On the other hand, for $PQ<0$, the  IAW packets are modulationally stable which may propagate in the form of  dark-envelope solitons characterized by a depression of the  wave potential around $\xi =0$. Typical form of this solution of Eq. \eqref{e39} is given by
\begin{equation}
\begin{split}
&\psi=\psi_{d0}~\text{tanh}^2\left(\frac{\xi-V\tau}{W_d}\right),   \\
&\theta=\frac{1}{2P}\left[V\xi-\left(\frac{V^2}{2}-2PQ\psi_{d0}\right)\tau\right].\label{dark-envelope}
\end{split}
\end{equation}
This represents a localized region of hole (void) traveling at a speed $V$. The pulse width $W_d$ depends on the constant amplitude $\psi_{d0}$ as $W_d=\sqrt{2|P/Q|/\psi_{d0}}$. Clearly, for a constant amplitude $\psi_{d0}$ the width of the dark envelope solitons have the similar properties as the bright solitons  with the variations of the plasma parameters. Typical profiles of the bright (at $k=0.7,~1.5$) and dark (at $k=0.1$) envelope solitons are shown in Fig. \ref{fig3} for the same parameter values as  for the dotted line in the lower panel of Fig. \ref{fig1}. It is seen that while  the bright envelope solitons exist at small and medium scales of perturbations, the dark envelope solitons can exist only at the large scale $k\ll1$.
\par
In the ranges of values of $k$ for which the IAW packet becomes unstable $(PQ>0)$,  Eq. \eqref{e39} can admit highly energetic rogue wave solutions.  These   rogue waves or rogons, in which a significant amount of energy is concentrated   in  a relatively small area in  space and time, are  generated due to the MI of the   coherent  IAW packets in the limit of infinite wave modulation period. In particular, they significantly amplify the  carrier wave amplitudes, and hence increase the nonlinearity during the evolution of the  wave packets. Now, Eq. \eqref{e39} can be rewritten as
\begin{equation}
i\frac{\partial \phi}{\partial \tilde{\tau}}+\frac12\frac{\partial^2 \phi}{\partial \tilde{\xi}^2}+|\phi|^2\phi=0, \label{NLSE-nondim}
\end{equation}
where $\tilde{\tau}=Q\tau$ and $\tilde{\xi}=\sqrt{Q/2P}\xi$.
The   rogue wave solution of Eq. \eqref{NLSE-nondim}  that is located on a non-zero background and localized  in  both  space and time  can be obtained as \cite{wong13}
\begin{equation}
\phi_{n}(\tilde{\xi},\tilde{\tau})=\left[(-1)^n+\frac{G_n(\tilde{\xi},\tilde{\tau})+iH_n(\tilde{\xi},\tilde{\tau})}{D_n(\tilde{\xi},\tilde{\tau})} \right]\exp(i\tilde{\tau}), \label{rogue-solution-general}
\end{equation}
where $G_n,~H_n$ and $D_n~(\neq0)$ are some polynomial functions of $\tilde{\xi}$ and $\tilde{\tau}$, and $n=1,2,3,\cdots$, denotes the order of the solution. The first-order rogon solution is obtained by considering $n=1$  and it  corresponds to the Peregrine soliton \cite{peregrine83} in which $G_1=4$, $H_1=8\tilde{\tau}$ and $D_1=1+4\tilde{\xi}^2+4\tilde{\tau}^2$. However, superposition of two first-order rogue waves can lead to the generation of another highly energetic rogue waves with higher amplitudes. The analytic form of these  waves have  been recently obtained by \textcite{akhmediev09} using the deformed Darboux transformation approach. This second order $(n=2)$ rogon solution has the same form as Eq. \eqref{rogue-solution-general} where the polynomials $G_2$, $H_2$ and $D_2$ are given by
\begin{equation}
G_2=-\left(\tilde{\xi}^2+\tilde{\tau}^2+\frac{3}{4}\right)\left(\tilde{\xi}^2+5\tilde{\tau}^2+\frac{3}{4}\right)+\frac{3}{4},
\end{equation}
\begin{equation}
H_2=\tilde{\tau}\left[3\tilde{\xi}^2-\tilde{\tau}^2-2\left(\tilde{\xi}^2+\tilde{\tau}^2\right)^2-\frac{15}{8}\right],
\end{equation}
\begin{equation}
\begin{split}
D_2=&\frac{1}{3}\left(\tilde{\xi}^2+\tilde{\tau}^2\right)^3+\frac{1}{4}\left(\tilde{\xi}^2-3\tilde{\tau}^2\right)^2\\
&+\frac{3}{64}\left(12\tilde{\xi}^2+44\tilde{\tau}^2+1\right).
\end{split}
\end{equation}
\par
\begin{figure}[ht]
\includegraphics[height=2.5in,width=3.3in]{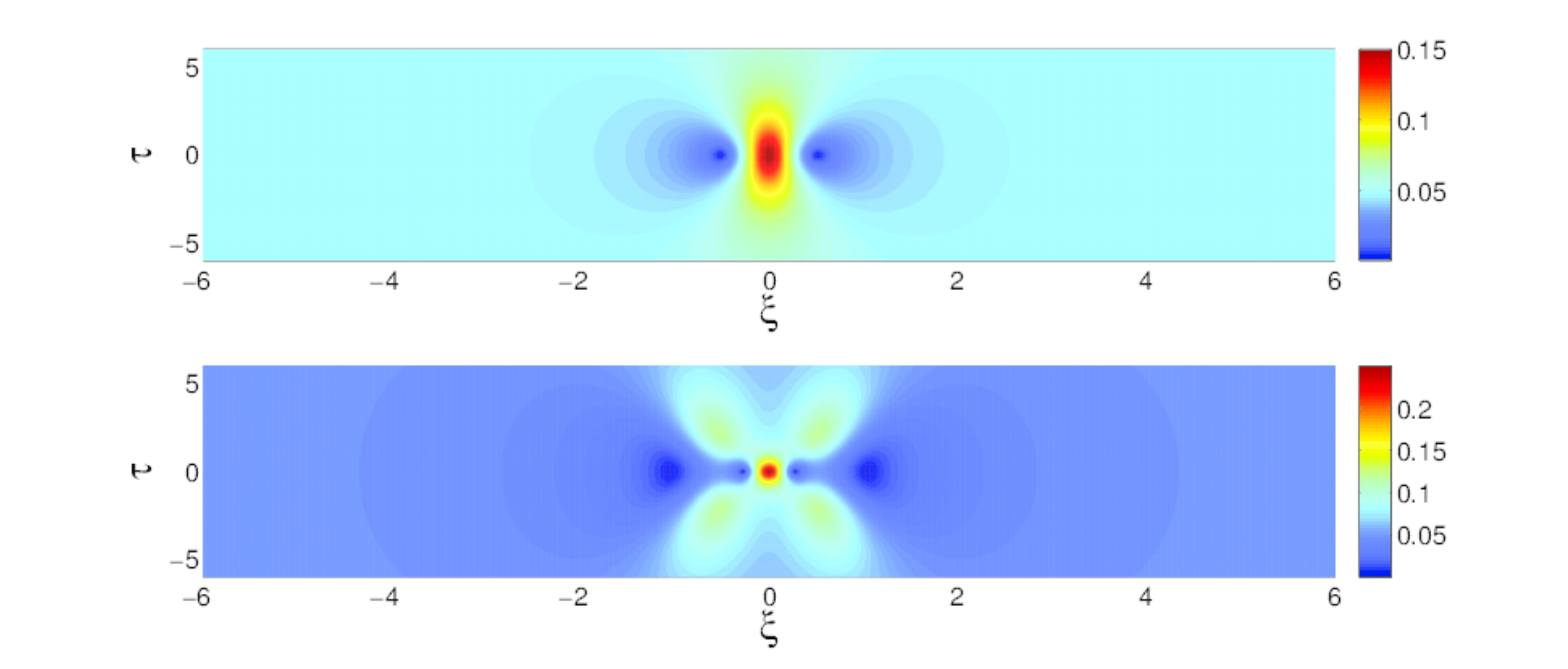}
\caption{Profiles (contour plots) of the first $(n=1)$ and second $(n=2)$ order rogon solutions given by Eq. \eqref{rogue-solution-general} of the NLS equation \eqref{NLSE-nondim} are shown in the upper and lower panels respectively. The parameter values are $q_c=0.4,~q_h=0.3,~f=0.2,~\beta=0.6$ and $k=1.5$.     }
\label{fig4}
\end{figure}
\begin{figure}[ht]
\includegraphics[height=2in,width=3.3in]{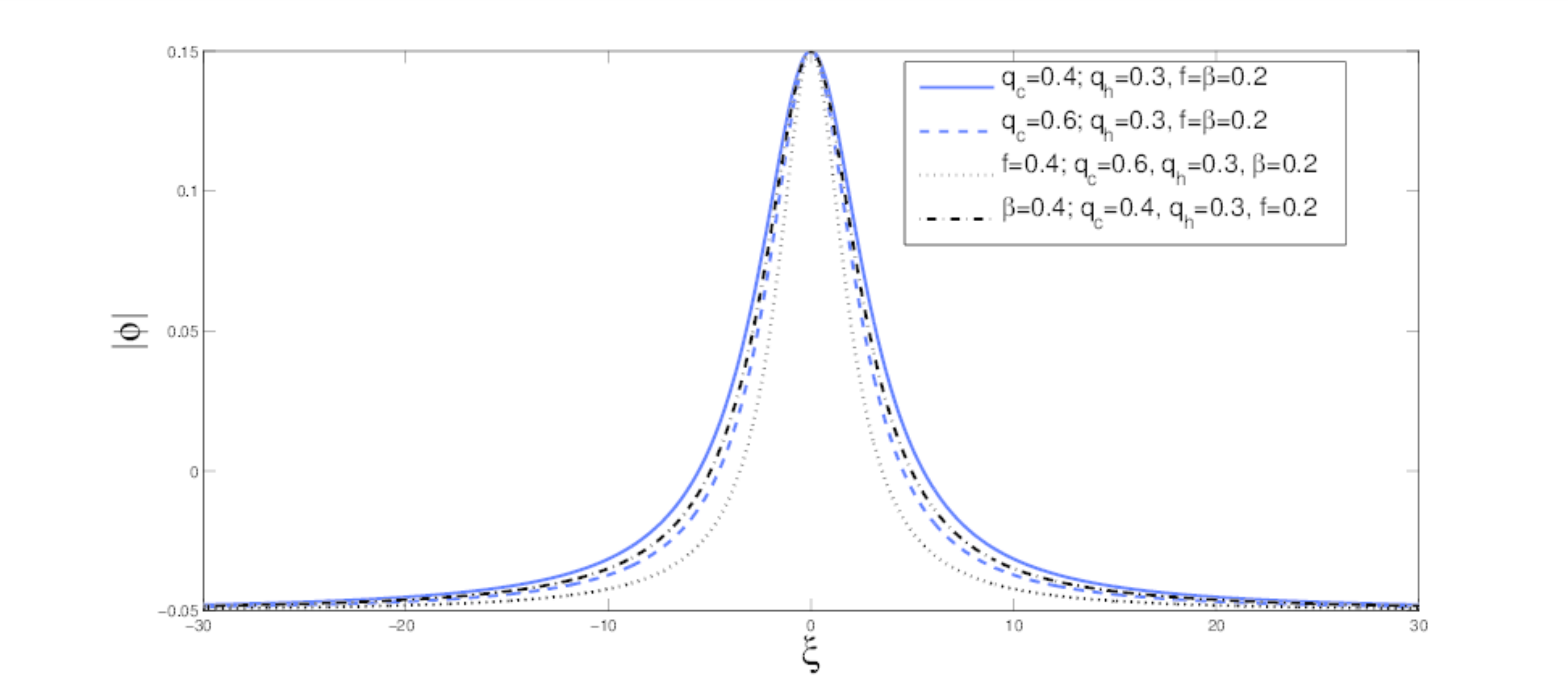}
\label{fig5a}
\includegraphics[height=2in,width=3.3in]{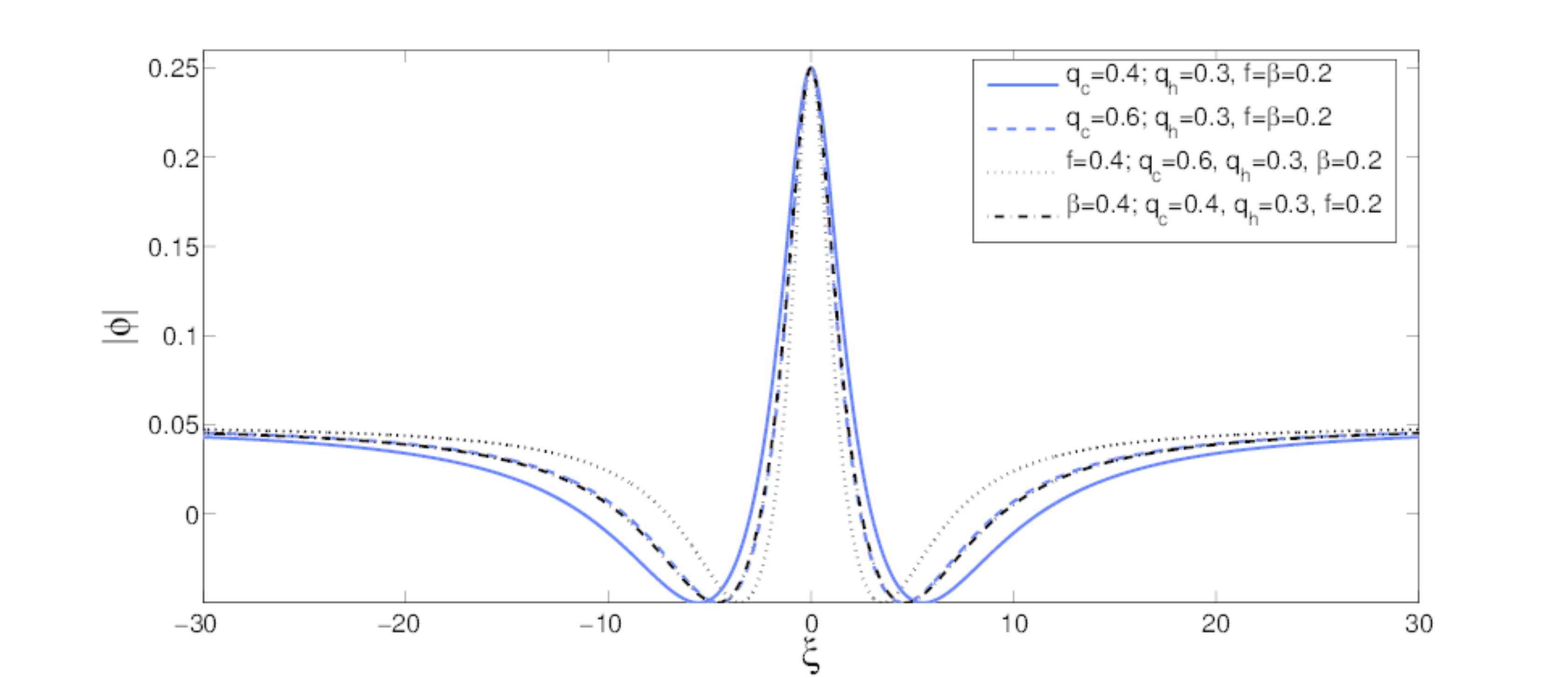}
\label{fig5b}
\caption{Properties of the first $(n=1)$ and second $(n=2)$ order rogon solutions at $\tau=0$ given by Eq. \eqref{rogue-solution-general} of the NLS equation \eqref{NLSE-nondim} are shown in the upper and lower panels respectively for different values of the parameters as in the figure. For the second order rogons  an increase of $q_c=0.6$ or $\beta=0.4$ effectively gives almost the same results (See the dashed and dash-dotted lines). }
\label{fig5}
\end{figure}

The profiles of these rogon solutions given by Eq. \eqref{rogue-solution-general} for $n=1,2$  are shown in Fig. \ref{fig4}. We find that for $k<1$ $(k>1)$, the localization occurs at large (small) $\xi$. The first-order Peregrine soliton has been recently observed experimentally in plasmas \cite{bailung11}. However, the second-order rogon solution, which has been observed in water waves \cite{chabchoub2012},
is yet to  observe in plasmas.  The  amplification factor of the amplitude of the $n$-th order rational solution [Eq. \eqref{rogue-solution-general}]  at $\tilde{\xi}=\tilde{\tau}=0$  is, in general, of $2n + 1$. Hence, localized  IAWs that are modeled by the higher-order breather solutions can also cause the formation of super rogue waves.
In order to  investigate the properties of these rogue wave solutions more clearly  for different values of the parameters $q_j,~f$ and $\beta$, we plot the solutions  at $\tau=0$. These are  shown in Fig. \ref{fig5}.  We find that while the amplitudes of both the first and second order rogons remain the same, their widths decrease with increasing values of $q_c$ (i.e., lack of superthermal cold electrons), $f$ and the temperature ratio $\beta$. However, no significant change of  the width  is found with a small increase of $\beta_h$ for both the rogons.
\section{Summary and conclusion}\label{VI}
In this work, we have investigated the amplitude modulation  of ion-acoustic wave (IAW) packets in an unmagnetized multi-component plasma consisting of singly charged positive ions and two-temperature electrons featuring nonextensive distributions. Using the multiple-scale perturbation method, a nonlinear Schr{\"o}dinger (NLS) equation is derived which governs the evolution of IAW envelopes. This equation is then used to study the modulational instability (MI) of a Stokes' wave train to a small longitudinal perturbation. Different stable and unstable regions of modulation in the range of the carrier wave number $k$ are found out  with the variations of the plasma parameters, namely, the nonextensive indices $q_c$ and $q_h$, the ratio of cool electron to ion number densities $f$ as well as the cool to hot electron temperature ratio $\beta$. It follows that the modulated IAW packets can propagate in the form of bright envelope solitons or highly energetic rogue waves (rogons) due to MI as well as dark envelope solitons where the MI is excluded. The growth rate of instability together with the exact analytical solutions of the NLS equation in the form of bright and dark envelope solitons as well as rogons are also obtained and their properties are examined numerically with the plasma parameters. The main results can be summarized as follows:
\begin{itemize}
\item The fluid model governing the dynamics of small-amplitude IAWs in multi-component electron-ion plasmas is valid for the plasma parameters satisfying $0<f,~\beta,~q_c,~q_h<1$, i.e., plasmas with more superthermal electrons [compared to the Maxwellian ones $(q\rightarrow1)$] and   lower concentration of cold electrons than the positive ions.
\item The group velocity dispersive coefficient $P$ of the NLS equation is always negative irrespective of the values of $k$ and the plasma parameters. However, the nonlinear coefficient $Q$ is always negative for $k\gtrsim1$ (and any values of the parameters with $0<f,~\beta,~q_c,~q_h<1$) and can become positive in the limit $k\rightarrow0$ as well as in a sub interval of $0<k<1$ depending on the plasma parameters. Thus, the modulated IAW  packets are always stable (unstable) with long-wavelength (short-wavelength) perturbations (carrier modes) giving rise the formation of dark envelope  solitons (bright envelope solitons or highly energetic rogons). The wave envelope can even be unstable at a medium scale of perturbation with $0<k\lesssim1$.
 \item  As   $\beta$ and $f$ increase, the value of $Q$ decreases (hence decreasing the maximum growth rate of instability) and the range  of values of $k$ for which $Q<0$ (hence the MI) increases. The same is true for increasing values of $q_c$, but decreasing values of $q_h$ except for some region with $k\rightarrow0$.
 \item As the parameters $\beta,~f$ and $q_c$ increase, the values of the critical wave number  $K_c$ below which the MI sets in  increase, however, the same decreases with increasing values of $\beta_h$. It turns out that the growth rate of instability can be suppressed in superthermal plasmas with higher temperature of hot electrons than the cool electrons, the lower concentration of cold electrons than ions and more (less) superthermal cool (hot) electrons.
\item  In the range of values of $k$ where $PQ>0$, bright envelope solitons are shown to exist both  at the small $(k>1)$ and medium $(k\lesssim1)$ scale of perturbations. However, dark solitons exist only at large scales $(k\ll1)$.
  \item  The MI of IAW envelopes can also lead to the generation of   rogue waves in which a significant amount of wave energy is concentrated in relatively a small area in space and time. Typical profiles of the first and second-order rogons are presented with respect to finite space and time. It is found that at an equilibrium state,  while the amplitudes of both the first and second order rogons remain the same, their widths decrease with increasing values of $q_c$ (i.e., lack of superthermal cold electrons), $f$ and the temperature ratio $\beta$. However, no significant change of  the width  is found with a small increase of $\beta_h$ for both the rogons.
 \end{itemize}
 To conclude, the findings of the present investigation should be relevant both in space and astrophysical plasma environments where two-temperature electrons are the main constituents together with ions. We also propose that an experiment should be done to observe the role of higher order effects for the formation of highly energetic ion-acoustic rogue waves (super rogue waves).\\

\textbf{Acknowledgment}\\
Shalini thanks University Grants Commission, New Delhi for awarding Rajiv-Gandhi Fellowship. The work by N.S.S. was supported by University Grants Commission, New Delhi, India under the major research project $F. No. 41-873/2012(SR)$.


\begin{thebibliography}{200}
\bibitem[Hershkowitz(1985)]{hershkowitz85}N. Hershkowitz, \textit{Space Sci. Rev.} \textbf{41}, 351 (1985).
\bibitem[Nishida and Nagasawa(1986)]{nishida86}Y. Nishida and T. Nagasawa, \textit{Phys. Fluids}, \textbf{29}, 345 (1986).
\bibitem[Hairapetian and Stenzel(1990)]{hairapetian90}G. Hairapetian and R. L. Stenzel, \textit{Phys. Rev. Lett. 65}, \textbf{175} (1990).
\bibitem[Temerin et al.(1982)]{temerin82}M. Temerin, K. Cerny, W. Lotko, F. S. Mozer: \textit{Phys. Rev. Lett.}, \textbf{48}, 1175 (1982).
\bibitem[Jones et al.(1975)]{jon75} W. D. Jones, A. Lee, S. N. Gleeman, H. J. Doucet, \textit{Phys. Rev. Lett.} \textbf{35}, 1349 (1975).
\bibitem[Goswami and Buti(1976)]{gos76}B. N. Goswami and B. Buti, \textit{Phys. Lett. A}, \textbf{57}, 149 (1976).
\bibitem[Nishihara and Tajiri(1981)]{nis81}K. Nishihara, M. Tajiri: \textit{Phys. Soc. Jpn}., \textbf{50}, 4047 (1981).
\bibitem[Sayal et al.(1993)]{say93}V. K. Sayal, L. L. Yadav, S. R. Sharma, \textit{Phys. scr.} \textbf{47}, 576 (1993).
\bibitem[Rice et al.(1993)]{ric93}W. K. M. Rice, M. A. Hellberg, R. Mace, S. Baboolal: \textit{Phys. Lett. A} \textbf{174}, 416 (1993).
\bibitem[Pottelette et al.(1999)]{pottelette99}R. Pottelette, R. E. Ergun, R. A. Treumann, M. Berthomier, C. W. Carlson, J. P. McFadden, I. Roth \emph{Geophys. Res. Lett.}, \textbf{26}, 2629(1999).
\bibitem[Ergun et al.(1998)]{ergun98}R. E. Ergun, C. W. Carlson, J.P. McFadden, F. S. Mozer, G. T. Delory, W. Peria, C. C. Chaston, M. Temerin, R. Elphic, R. Strangeway, R. Pfaff, C. A. Cattell, D. Klumpar, E. Shelley, W. Peterson, E. Moebius, L. Kistler: \textit{Geophys. Res. Lett.}, \textbf{25}, 2061 (1998).
\bibitem[Mcfadden(2003)]{mcfadden03}J. P. Mcfadden, et al.: \emph{J. Geophys. Res. }, \textbf{108}, 8018 (2003)
\bibitem[Baluku et al.(2012)]{bal12}T. K. Baluku and M. A. Hellberg, \textit{Phys. Plasmas}, \textbf{19}, 012106 (2012).
\bibitem[Washimi and Taniuti(1966)]{was66}H. Washimi, T. Taniuti, \textit{Phys. Rev. lett.} \textbf{17}, 996 (1966).
\bibitem[Tajiri and Nishihara(1984)]{tajiri84}M. Tajiri and K. Nishihara \textit{J. Phys. Soc. Japan} \textbf{54} 572 (1984); D. Chakraborty and K. P. Das \textit{Phys. Plasmas} \textbf{10} 2236 (2003).
\bibitem[Sagdeev(1966)]{sagdeev66}R. Z. Sagdeev \textit{Reviews of Plasma Physics} \textbf{4} 52 (1966).
\bibitem[Ikezi(1970)]{ikezi70}H. Ikezi \textit{Phys. Rev. Lett.} \textbf{25} 11 (1970); Y. Nakamura et al. \textit{J. Plasma Phys.} \textbf{33} 237 (1985); Y. Nakamura and I. Tsukabayashi \textit{J. Plasma Phys.} \textbf{34} 401 (1985).
\bibitem[Sittler et al.(1983)]{sit83}E. C. Jr. Sittler, K. W. Ogilvie, J. D. Scudder, \textit{J. Geophys. Res.} \textbf{88}, 8847 (1983).
\bibitem[Barbosa and Kurth(1993)]{bab93}D. D. Barbosa and W. S. Kurth, \textit{J. Geophys. Res.} \textbf{98}, 9351 (1993).
\bibitem[Renyi(1955)]{ren55}A. Renyi: \textit{Acta Math. Acad. Sci. Hung.}, \textbf{6}, 285 (1955).
\bibitem[Tsallis(1988)]{tsa88}C. Tsallis, \textit{J. Stat. Phys.}, \textbf{52}, 479 (1988).
\bibitem[Caruso and Tsallis(2008)]{car08}F. Caruso and C. Tsallis, \textit{Phys. Rev. E} \textbf{78}, 021102 (2008).
\bibitem[Tribeche and Djebarni(2010)]{tri10}M. Tribeche, L. Djebarni: \textit{Phys. Plasmas}, \textbf{17}, 124502 (2010).
\bibitem[Nobre et al.(2011)]{nob11}F. D. Nobre, M. A. Rego-Monteiro, C. Tsallis, \textit{Phys. Rev. Lett.} \textbf{106}, 140601 (2011).
\bibitem[Bains et al.(2011)]{bai11}A. S. Bains, M. Tribeche and T. S. Gill, \textit{Phys. Plasmas}, \textbf{18}, 022108, (2011).
\bibitem[Tribeche and  Merriche(2011)]{tri11}M. Tribeche, A. Merriche: \textit{Phys. Plasmas}, \textbf{18}, 034502 (2011).
\bibitem[Sahu and Tribeche(2012)]{sah12}B. Sahu, M. Tribeche, \textit{Astrophys. space sci.} \textbf{341}, 573 (2012).
\bibitem[El-Taibany and Tribeche(2012)]{tai12}W. F. El-Taibany, M. Tribeche: \textit{Phys. Plasmas}, \textbf{19}, 024507 (2012).
\bibitem[Saini and Kourakis (2008)]{saini08}N. S. Saini and I. Kourakis, \textit{Phys. Plasmas}, \textbf{15}, 123701 (2008).
\bibitem[Kourakis and Shukla(2003)]{kourakis03}I. Kourakis and P. K. Shukla, \textit{J. Phys. A: Math. Gen.} \textbf{36}, 11901 (2003).
\bibitem[Sultana et al.(2014)]{sultana14}S. Sultana, S. Islam and A. A. Mamun, \textit{AstroPhys. Space Sci.} \textit{351} 581-589 (2014).
\bibitem[Kharif et al.(2009)]{kharif09}C. Kharif, E. Pelinovsky, and  A. Slunyaev 2009 \textit{Rogue Waves in the Ocean (Springer-Verlag, Berlin}.
\bibitem[Eliasson and Shukla(2010)]{eliasson10}B. Eliasson and P. K. Shukla, \textit{Phys. Rev. Lett.}, \textbf{105}, 014501 (2010).
\bibitem[Yan(2010)]{yan10}Z. Yan, \textit{Commun. Theor. Phys.}, \textbf{54}, 947 (2010).
\bibitem[Montina et al.(2009)]{montina09}A. Montina, U. Bortolozzo, S. Residori, and F. T. Arecchi, \textit{Phys.Rev. Lett.}, \textbf{103}, 173901 (2009).
\bibitem[Kibler et al.(2010)]{kibler10}B. Kibler, J. Fatome, C. Finot, G. Millot, F. Dias, G. Genty, N. Akhmediev and J. M. Dudley, \textit{Nature Phys. (London)}, \textbf{6}, 790 - 795 (2010).
\bibitem[Ganshin et al.(2008)]{ganshin08}A. N. Ganshin, V. B. Efimov, G. V. Kolmakov, L. P. Mezhov-Deglin and P. V. E. McClintock, \textit{Phys. Rev. Lett.}, \textbf{101}, 065303 (2008).
\bibitem[Bludov et al.(2009)]{bludov09}Yu. V. Bludov, V. V. Konotop and N. Akhmediev, \textit{Phys. Rev. A}, \textbf{80}, 033610 (2009).
\bibitem[Stenflo and Marklund(2010)]{stenflo10}L. Stenflo and M. Marklund, \textit{J. Plasma Phys.}, \textbf{76}, 293 - 295 (2010).
\bibitem[Onorato(2011)]{onorato11}M. Onorato, D. Proment, and A. Toffoli, \textit{Phys. Rev. Lett.}, \textbf{107}, 184502 (2011).
\bibitem[Shukla et al.(2006)]{shukla06}P. K. Shukla, I. Kourakis, B. Eliasson, M. Marklund, and L. Stenflo \textit{Phys. Rev. Lett.} \textbf{97} 094501 (2006).
\bibitem[Peregrine(1983)]{peregrine83}D. H. Peregrine, \textit{J. Aust. Math. Soc. Ser. B Appl. Math.}, \textbf{25}, 16-43 (1983).
\bibitem[Dysthe and Trulsen(1999)]{dysthe9}K. B. Dysthe, K. Trulsen, \textit{Phys. Scr.}, \textbf{T82}, 48 - 52, (1999).
\bibitem[Chabchoub et al.(2011)]{chabchoub11}A. Chabchoub, N. P. Hoffmann and N. Akhmediev, \textit{Phys. Rev. Lett.}, \textbf{106}, 204502 (2011).
\bibitem[Bailung  et al.(2011)]{bailung11}H. Bailung, S. K. Sharma and Y. Nakamura \textit{Phys. Rev. Lett.}, \textbf{107}, 255005 (2011).
\bibitem[Clamond et al.(2006)]{clamond06}D. Clamond, M. Francius, J. Grue, C. Kharif, \textit{Eur. J. Mech. B (Fluids)}, \textbf{25}, 536 - 553 (2006).
\bibitem[Akhmediev et al.(2009)]{akhmediev09}N. Akhmediev, A. Ankiewicz and M. Taki, \textit{Phys. Lett. A.}, \textbf{373}, 675 (2009).
\bibitem[Zhong et al. (2013)]{wong13}W. P. Zhong, M. R. Belic and T. Huang, \textit{Phys. Rev. E} \textbf{87}, 065201 (2013).
\bibitem[Voronovich et al.(2008)]{vor08}V. V. Voronovich, V. I. Shrira and G. Thomas \textit{J. Fluid Mech.}, \textbf{604}, 263 - 296 (2008).
\bibitem[Ruderman(2010)]{ruderman10}M. S. Ruderman, \textit{Eur. Phys. J. Special Topics}, \textbf{185}, 57 - 66 (2010).
\bibitem[Shats et al.(2010)]{shats10}M. Shats, H. Punzmann and H. Xia, \textit{Phys.Rev. Lett.}, \textbf{104}, 104503 (2010).
\bibitem[El-Awady and Moslem(2011)]{awady11}E. I. El-Awady and W. M. Moslem, \textit{Phys. Plasmas}, \textbf{18}, 082306 (2011).
\bibitem[Abdelsalam et al.(2011)]{abdelsalam11}U. M. Abdelsalam, W. M. Moslem, A. H. Khater and P. K. Shukla, \textit{Phys. Plasmas}, \textbf{18}, 0923051 (2011).
\bibitem[Moslem et al.(2011)]{moslem11}W. M. Moslem, R. Sabry, S. K. El-labany and P. K. Shukla, \textit{Phys. Rev. E.}, \textbf{84} 066402 (2011).
\bibitem[Shukla and Moslem(2012)]{shukla12}P.K. Shukla and W. M. Moslem, \textit{Phys. Lett. A.}, \textbf{376}, 1125 (2012)
\bibitem[Bains et al.(2014)]{bains14}A. S. Bains, Bo Li and Li-Dong Xia, \textit{Phys. Plasmas}, \textbf{21}, 032123 (2014).
\bibitem[Veldes et al.(2013)]{veldes13}G. P. Veldes, J. Borhanian, M. McKerr, V. Saxena, D. J. Frantzeskakis and I. Kourakis, \textit{J. Opt.},  \textbf{15}, 064003 (2013).
\bibitem[Abdelsalam(2012)]{abdelsalam12}U. M. Abdelsalam, \textit{J. Plasma Phys.}, \textbf{79}, 287 � 294 (Cambridge University Press 2012), (2012)\\ \textit{doi:10.1017/S0022377812000992}.
\bibitem[Guo et al.(2013)]{guo13}Guo Shimin, Mei Liquan and Shi Weijuan, \textit{Phys. Lett. A}, \textbf{377}, 2118 � 2125 (2013).
\bibitem[El-Wakil et al.(2014)]{el-wakil14}S. A. El-Wakil, Abulwafa M. Essam, A. El-hanbaly and E. K. El-Shewy  \textit{Astrophys. Space Sci.}, \textbf{353}, 501 � 506 (2014).
\bibitem[Shalini and Saini(2015)]{shalini15}Shalini and N. S. Saini, \textit{J. Plasma Phys.}\\
 \textit{doi:10.1017/S0022377815000082} (2015).
\bibitem[Asano et al.(1969)]{asa69}N. Asano, T. Taniuti and N. Yajima, \textit{Astrophys. J.} \textbf{10}, 2020 (1969).
\bibitem[Kourakis and Shukla(2005)]{kou05}I. Kourakis and P. K. Shukla, \textit{Nonlinear Proc. Geophys.} \textbf{12}, 407 (2005).
\bibitem[Dauxois and Peyrard(2006)]{dauxois06}T. Dauxois and M. Peyrard, \textit{Physics of Solitons}, (Cambridge: Cambridge University Press) (2006).
\bibitem[Hasegawa(1975)]{has75}A. Hasegawa, \textit{Plasma Instabilities and Nonlinear Effects} (1975).
\bibitem[Sultana and Kourakis(2011)]{sultana11}S. Sultana and I. Kourakis, \textit{Plasma Phys. Control. Fusion} \textit{53} 045003 (2011).
\bibitem[Fedele et al.(2002)]{fedel02}R. Fedele, H. Schamel and P. K. Shukla, \textit{Phys. Scr.} \textbf{T98}, 18 (2002).
\bibitem[Fedele and Schamel(2002)]{fed02}R. Fedele and H. Schamel, \textit{Eur. Phys. J.} \textbf{B27}, 313 (2002).
\bibitem[Chabchoub et al.(2012b)]{chabchoub2012}A. Chabchoub, N. P. Hoffmann, M. Onorato and N. Akhmediev,  \textit{Phys. Rev. X}, \textbf{2}, 011015, (2012).

\end{thebibliography}
\end{document}